# Nonlinear dynamical properties of an oscillating tip–cantilever system in the tapping mode

L. Nony, R. Boisgard, and J. P. Aimé[a)]
*CPMOH Université Bordeaux I, 351 Cours de la Libération, F-33405 Talence, France*

The dynamical properties of an oscillating tip–cantilever system are now widely used in the field of scanning force microscopy. The aim of the present work is to get analytical expressions describing the nonlinear dynamical properties of the oscillator in noncontact and intermittent contact situations in the tapping mode. Three situations are investigated: the pure attractive interaction, the pure repulsive interaction, and a mixing of the two. The analytical solutions obtained allow general trends to be extracted: the noncontact and the intermittent contact show a very discriminate variation of the phase. Therefore the measurement of the phase becomes a simple way to identify whether or not the tip touches the surface during the oscillating period. It is also found that the key parameter governing the structure of the dynamical properties is the product of the quality factor by a reduced stiffness. In the attractive regime, the reduced stiffness is the ratio of an attractive effective stiffness and the cantilever one. In the repulsive regime, the reduced stiffness is the ratio between the contact stiffness and the cantilever one. The quality factor plays an important role. For large values of the quality factor; it is predicted that a pure topography can be obtained whatever the value of the contact stiffness. For a smaller quality factor, the oscillator becomes more sensitive to change of the local mechanical properties. As a direct consequence, varying the quality factor, for example with a vacuum chamber, would be a very interesting way to investigate soft materials either to access topographic information or nanomechanical properties.

## I. INTRODUCTION

Recent developments in the field of scanning force microscopy (SFM) are focused on the behavior of the dynamical properties of an oscillating tip–cantilever system (OTCL). Experimentally, the use of the OTCL to probe surface properties at the local scale, from the nanometer to a few picometers, is done with two different operating modes.

One mode keeps the oscillating amplitude constant, and the measured quantities are shift in resonance frequency as a function of the tip–surface distance. In this case, recording image is obtained by moving up and down the surface in order to keep a chosen frequency shift constant. These experiments are performed without any contact between the tip and the sample; thus frequency shifts are a measure of the attractive force between the tip and the surface. Spectacular experimental results have shown that images giving a contrast at a few ten picometers can be achieved.[1–5] Companion theoretical developments suggest that the high sensitivity of this mode is due to the nonlinear dynamical properties of the OTCL at the proximity of the surface.[6–8] Recent experimental results show variations of the image as a function of the chosen frequency shift.[4] Here again, it does appear that the images obtained depend of several factors, among them the OTCL properties and the fine structure of the tip. There remains a quite long way to completely understand the image obtained, a general and well-known problem in the field of the local probe method.

With the second mode, a drive frequency is chosen. The feedback loop is used to maintain constant the amplitude of the OTCL. The recorded images are the vertical displacements needed to keep the amplitude constant and the corresponding phase of the OTCL. This mode, commonly called tapping, is often used in intermittent contact (IC), that is, during a part of the oscillating period the tip touches the surface but images can also be recorded without any contact. This mode had been conceived mainly to reduce the shear forces at the interface between the tip and the surface. So that a new area is open in which soft materials, polymers, and biological systems can be investigated without producing significant damages. Numerous experimental results have shown the ability of this mode to image soft materials.[9–12] Among them, phase contrast inversion as a function of the set point and images of triblock copolymers are quite convincing of the great potentiality of this mode. But here again, recording a true topography is far to be achieved; more correctly one may discuss in terms of nanomechanical properties of the sample probed by this mode.[12–16] Several theoretical approaches have been dedicated to the tapping, part of them being numerical simulations.[14,17–21] The major drawback of numerical simulations is the difficulty to give general routes and simple predictions to understand and define experimental strategies.

The present work is an attempt to extract analytical expressions describing the tapping mode. As shown in preceding papers,[7,19] the nonlinear dynamics of the OTCL in an attractive field is quite complicated and indeed do show a subtle different behavior depending of the mode chosen.[7,14,20,22] Even more complicated is the situation when

[a)]Electronic mail: jpaime@cribx1.u-bordeaux.fr

the tip is brought into contact to the sample. Therefore very simple hypotheses are used in order to extract analytical solutions giving the essential trends of the OTCL response.

The paper is organized as follows: the theoretical part is divided in two sections. The attractive regime is presented in Sec. II. The phase and amplitude variations as a function of the OTCL-surface distance are shown for a drive frequency slightly below the resonance frequency. In Sec. III, the IC case is first discussed with a simple repulsive harmonic potential, then an attractive interaction is added. Section IV is dedicated to a brief discussion and a few experimental results are given. In Appendix A and B are given a detailed technical description of the theoretical approach and of the experimental conditions, respectively.

## II. ATTRACTIVE REGIME

In a tapping experiment, a tip–microlever system is kept vibrating at a drive frequency $\omega$ near a surface. Recording of the phase and vertical displacements of the piezoelectric ceramic holding the sample are made thanks to a feedback loop keeping the amplitude constant. To understand the origin of the image contrast, one must understand changes of the amplitude and phase, $A(D)$ and $\varphi(D)$, respectively, as functions of the OTCL-surface distance $D$. Recording $A(D)$ and $\varphi(D)$ is achieved by making approach–retract curves. The scan is stopped at a given location $(X;Y)$ in the horizontal plane of the sample, then a periodic motion along the vertical $Z$-axis is performed. In these experiments, series of approach–retract curves are obtained providing information on the properties of the dynamical behavior of the OTCL as a function of $D$ and properties of the sample at the local scale. The ultimate goal is to extract the sample properties from variations of $A(D)$ and $\varphi(D)$.

The attractive interaction between the tip and the sample is assumed to be a sphere–plane interaction involving the disperse part of the van der Waals interaction [see Eq. (5)]. For distances from the surface above a few nanometers, the attractive force is negligible and a forced, damped oscillator satisfactorily describes the OTCL. The differential equation describing the harmonic behavior, i.e., amplitude and phase relationships with the frequency, is given by:

$$\ddot{x}(t)+\frac{\omega_0}{Q}\dot{x}(t)+\omega_0^2 x(t)=\frac{f}{m}\cos(\omega t), \tag{1}$$

with, respectively, $\omega_0$, $Q$, and $m$ the resonance frequency, quality factor, and effective mass of the OTCL. $f$ and $\omega$ are the external drive force and drive frequency.
The stationary solution is:

$$x(t)=A_{\text{free}}(\omega)\cos[\omega t+\varphi_{\text{free}}(\omega)], \tag{2}$$

where $A_{\text{free}}(\omega)$, and $\varphi_{\text{free}}(\omega)$ are the amplitude and phase of the oscillations at the drive frequency. Since the oscillator responds with delay to the excitation, the sign of the phase chosen in Eq. (2) means that $\varphi_{\text{free}}(\omega)$ varies in the domain $[-180°;\ 0°]$. With reduced coordinates: $u=\omega/\omega_0$ and $a_{\text{free}}(u)=A_{\text{free}}(u)/A_0$, with $A_0$ the amplitude at the resonance frequency $(u=1)$ given by:

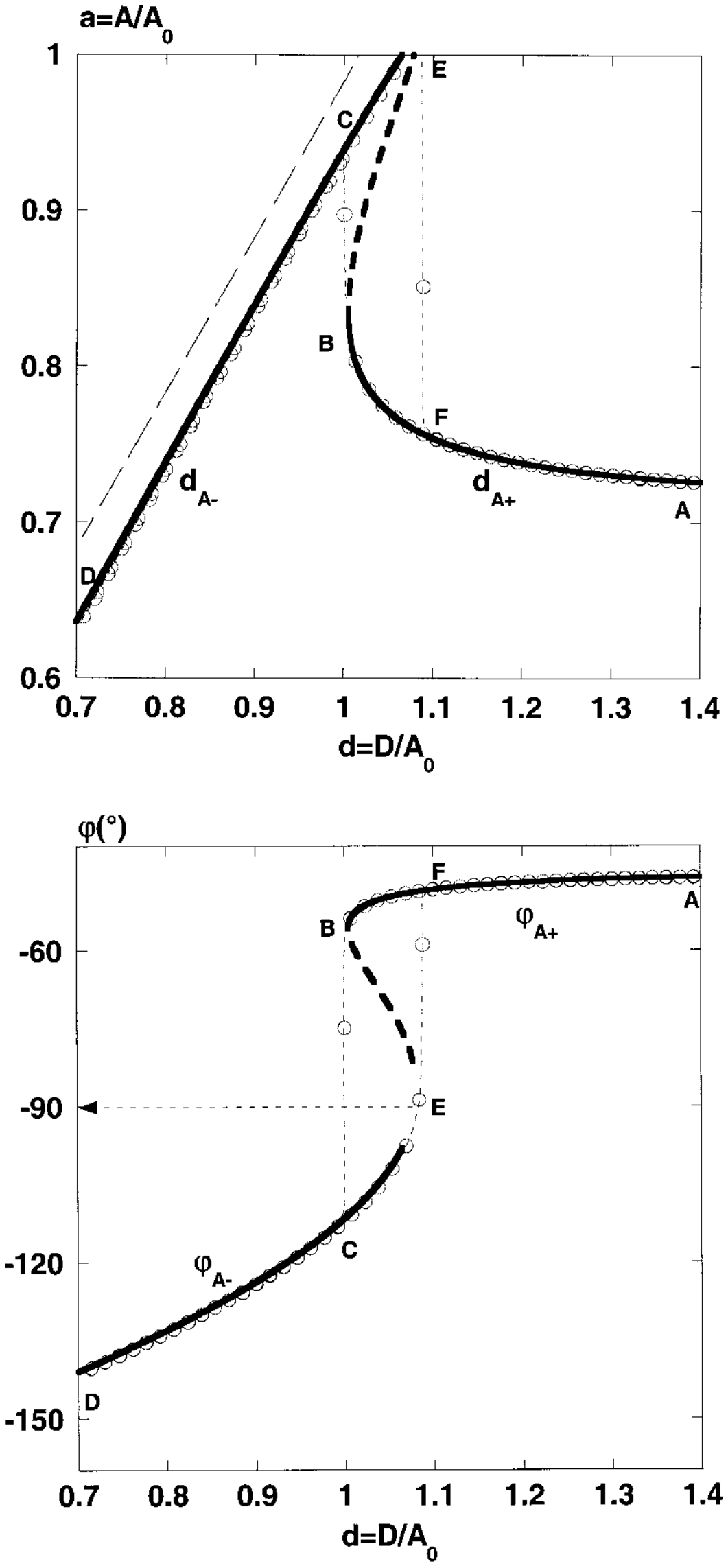

FIG. 1. (a) Variations of the amplitude as a function of the OTCL-surface distance in a pure attractive field corresponding to an approach–retract curve. Numerical simulation (empty circles) and calculated with Eq. (7a) (thick continuous line). The unstable part of the analytical branch $d_{A+}$ (point B to E) is represented with a thick dotted line. The parameters used are $A_0=10$ nm, $Q=50$, $u=0.990\,05$, $k_c=4.935$ N m$^{-1}$, $H=8.1675\times10^{-19}$ J, $R=20$ nm. The corresponding value of $\kappa_a$ is $\kappa_a=3.31\times10^{-3}$ $(Q\kappa_a=0.1655)$. The free amplitude is $a_{\text{free}}=1/\sqrt{2}=0.707$. The equation $d=a$ (thin dashed line) gives the location of the surface. The bistable structure is shown with the cycle BCEF. (b) Variations of the phase associated to the variation of the amplitude given in Fig. 1(a) [Eq. (7b)]. The phase jump leads to a noncontact phase below $-90°$. The analytical branches $\varphi_{A+}$ and $\varphi_{A-}$ don't join exactly. This is uniquely due to the numerical sampling used with the software (Kaleidagraph, r3.09).ny, Boisgard, and Aimé

$$A_0=\frac{f}{m}\frac{Q}{\omega_0^2}, \tag{3}$$

the solutions of Eq. (1) are:

$$a_{\text{free}}(u)=\frac{1}{\sqrt{Q^2(1-u^2)^2+u^2}} \tag{4a}$$

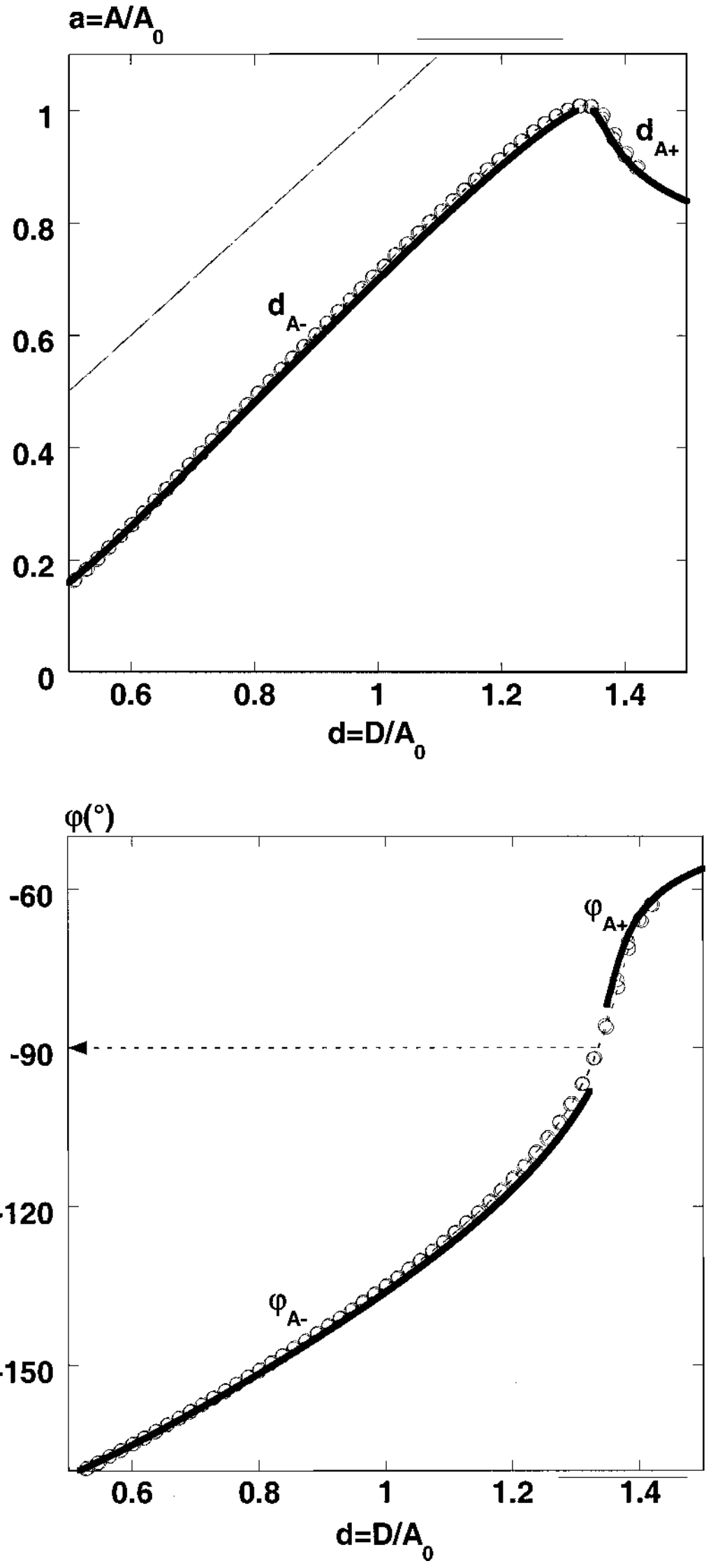

FIG. 2. (a) Variation of the amplitude similar to the one shown in Fig. 1(a) but with the parameters: $A_0 = 6$ nm, $Q = 50$, $u = 0.990\,05$, $k_c = 0.247$ N m$^{-1}$, $H = 10^{-19}$ J, $R = 20$ nm, corresponding to a larger value of $\kappa_a$, $\kappa_a = 3.752 \times 10^{-2}$ ($Q\kappa_a = 1.876$). The unstable behavior and the hysteresis cycle have disappeared. (b) Variation of the phase associated to the variation of the amplitude shown in Fig. 2(a) [Eq. (7b)]. The phase jump disappears and the phase crosses continuously the $-90°$ value, the noncontact phase is below $-90°$.onlinear dynamics of an oscillating tip-cantilever

$$\varphi_{\mathrm{free}}(u) = \arctan\left( \frac{u}{Q(u^2-1)} \right). \tag{4b}$$

During an approach–retract curve, the oscillator experiences both a pure attractive and an attractive–repulsive field. A van der Waals attractive field leads to a nonlinear dynamic behavior of the OTCL and can generate a hysteresis cycle in the approach–retract curve.[16–19] An approach based on a variational principle had shown that analytical expressions could be derived describing the nonlinear behavior of the OTCL.[19]

The aim of parts II and III is to develop the analytical model with a more precise description of the noncontact (NC) and IC situations including phase variations during an approach–retract curve. In Appendix A, detailed calculations are given. The main results are presented below.

In this part, we assume that the OTCL never touches the surface and is described as a forced damped oscillator plus an interacting van der Waals disperse term with a sphere–plane geometry.[23] The dissipation function added to the Lagrangian is $W = xF_d$, where $F_d = -(m\omega_0/Q)\underline{\dot{x}}$ is calculated along the physical path, thus $\overline{\dot{x}}$ is not a varied parameter.[24]

$$L = T - U + W \tag{5}$$

$$L = \frac{1}{2} m\dot{x}^2(t) - \left[ \frac{1}{2} m\omega_0^2 x^2(t) - x(t) f \cos(\omega t) - \frac{HR}{6(D - x(t))} \right] - \frac{m\omega_0}{Q} x(t) \underline{\dot{x}(t)},$$

where $H$ is the Hamaker constant, $R$ the tip's apex radius and $D$ the distance between the surface of the sample and the equilibrium position at rest of the OTCL. We focus on the stationary harmonic state, thus the trial function used has the stationary harmonic form of Eq. (2). With the reduced coordinate $d = D/A_0$, the analytical approach leads to the two coupled equations:

$$\cos(\varphi) = Qa(1-u^2) - \frac{aQ\kappa_a}{3(d^2-a^2)^{3/2}}$$
$$\sin(\varphi) = -ua, \tag{6}$$

where $\kappa_a = HR/k_c A_0^3$ is a dimensionless parameter. Since $k_c = m\omega_0^2$ is the cantilever stiffness, $HR/A_0^3$ has the dimension of a stiffness and can be related to a strength of the attractive interaction. Therefore varying $\kappa_a$ with $A_0$ is equivalent to change the strength of the attractive interaction: for example, a large (small) $A_0$ corresponds to a small (large) $\kappa_a$. The solutions of Eqs. (6) are:

$$d_{A\pm} = \sqrt{a^2 + \left( \frac{Q\kappa_a}{3\left( Q(1-u^2) \mp \sqrt{\frac{1}{a^2} - u^2} \right)} \right)^{2/3}} \tag{7a}$$

$$\varphi_{A\pm} = \arctan\left( \frac{u}{Q(u^2-1) + \frac{Q\kappa_a}{3(d_{A\pm}^2 - a^2)^{3/2}}} \right). \tag{7b}$$

Note that setting $\kappa_a = 0$ in Eq. (7b), i.e., no attractive interaction, leads to the phase equation of the harmonic oscillator [Eq. (4)]. As a consequence of the nonlinear behavior, at a given distance $d$, the amplitude and phase of the OTCL depends on the drive amplitude through the cubic term $A_0^3$.

Equations (7a) and (7b) give two physical branches for the amplitude and the phase (see Appendix A). For drive frequencies slightly below the resonance one amplitude and phase exhibit a hysteresis cycle that disappears as the magnitude of $Q\kappa_a$ increases [see Figs. 1(a), 1(b), and 2(a), 2(b)].

In Figs. 1 and 2 are reported amplitude and phase variations calculated from Eqs. (7a) and (7b) and the correspond-ing numerical results. The numerical simulation solves the following differential equation based on Eq. (1) plus the interacting term, $\ddot{x}(t)+(\omega_0/Q)x(t)+\omega_0^2 x(t)=(f/m)\cos(\omega t)-HR/6m(D-x(t))^2$, with a Runge–Kutta 4 method from arbitrary initial conditions. We assume that the oscillator is in its harmonic state for each step of the implementation. This is validated by an adiabatic criterion, which evaluates the number of points for the whole calculus for a given value of the quality factor. Amplitude and phase variations are then calculated thanks to a numerical synchronic detection filtering the first harmonic, $n=1$. To verify that the use of the harmonic solution was efficient to describe the nonlinear behavior, thus validate the analytical model, several numerical simulations were done with a synchronic detection filtering the second harmonic, $n=2$. We found that the amplitude variations were less than 0.3% of the ones for $n=1$. For the smaller $Q\kappa_a$ value [Figs. 1(a), 1(b)], letters illustrate the hysteresis cycle and the bistable behavior of the OTCL. When bifurcation occurs during the approach (retract), point B (E), the numerical solution jumps to its stable upper (lower) branch in amplitude, point C (F), instead of following the unstable branch BE.

Then, since $d$ decreases, the strength of the attractive field increases, from which an increase of the amplitude is expected. The reduction of the amplitude predicted, and observed,[16,19] is a nonintuitive result (C to D). The reduction of the amplitude is due to change of the phase relationship between the OTCL and the excitation, the phase getting a value below $-90°$ [Fig. 1(b) from C to D]. Thus, the attractive interaction acts as a repulsive one, leading to a decrease of the amplitude.[19]

The whole variation of the phase, phase jump, and cycle of hysteresis [Fig. 1(b)], is similar to the one of the amplitude [Fig. 1(a)]. For a small value of $Q\kappa_a$, the phase jumps below $-90°$ and an hysteresis cycle follows. For a large $Q\kappa_a$ value [Figs. 2(a), 2(b)], the phase crosses the value of $-90°$ continuously and the approach and retract curves are identical. Measurement of the phase is very useful to identify NC situations. As discussed below, as soon as the phase is less than $-90°$, the attractive regime is dominant.

At this stage, it is worth discussing in more detail the physical origin of the variation of the phase. Within the linear response theory, the instantaneous dissipated power is a function of the phase value through a sin function. When a nonlinear dynamical behavior occurs, the variation of the phase is not uniquely related to the dissipation but also to the distortion of the resonance peak.

As shown with the introduction of the dimensionless parameter $\kappa_a$ [Eqs. (7)], contrary to a linear behavior, the forcing term cannot be scaled out. The magnitude of the drive amplitude becomes a new operative parameter as much as the strength of the nonlinear coupling term. At fixed drive frequency and fixed drive amplitude, when the OTCL approaches the surface, the oscillating amplitude and the strength of the nonlinear coupling term vary. As a consequence, a phase variation is expected without involving a particular change of the dissipating process. In the present case, uniquely the attractive field is involved. From Eqs. (7) we derived the equation giving the shape of the resonance peak as a function of the OTCL-surface distance, $d$:

$$u=\sqrt{\frac{1}{a^2}-\left[\frac{1}{2Q}\left(1\pm\sqrt{1-4Q^2\left(1-\frac{1}{a^2}-\frac{\kappa_a}{3(d^2-a^2)^{3/2}}\right)}\right)\right]^2}. \tag{8}$$

Here we focus on the variations of the phase when a drive frequency is slightly below the resonance one. The case of a drive frequency at the resonance one or above gives a simpler structure and can be straightforwardly deduced. In Figs. 3(a) and 3(b) are given the evolutions of the resonance peak as a function of $d$ and the corresponding phase variations when the OTCL moves towards the surface. At distances $d$ large enough, say a few nanometers, the peak keeps its resonance shape and the phase remains roughly constant [domain (a)]. When the peak starts to distort and the location of $u$ crosses the bifurcation point, the phase jumps to the lower branch [domain (b)]. At a closer distance, the peak further distorts and the phase goes towards the $-180°$ value [domain (c)]. Therefore, the phase variation provides a precise information about the stationary state of the OTCL. This statement is particularly verified when IC situations take place.

## III. INTERMITTENT CONTACT (IC) REGIME

### A. Pure repulsive field

The IC situation is given by $a>d$. In a first step, the attractive interaction is considered as being negligible ($H=0$). The repulsive field acts during a short part of the OTCL period, typically a hundredth of the period. Thus the action is defined between the two instants $t_i=0$ and $t_f=[\arccos(D/A)]/\omega$. In order to get an analytical expression, the repulsive interaction is assumed to have a simple harmonic form: $(1/2)k_s(x(t)-D)^2$, where $k_s$ is the contact stiffness scaling as $k_s=G_s\phi$, with $\phi$ the diameter of the contact area involved in the IC situations and $G_s$ the Young's modulus of the sample.[25,26] For small indentations, $(a-d)/a\ll 1$, the variational principle gives the couple of equations:

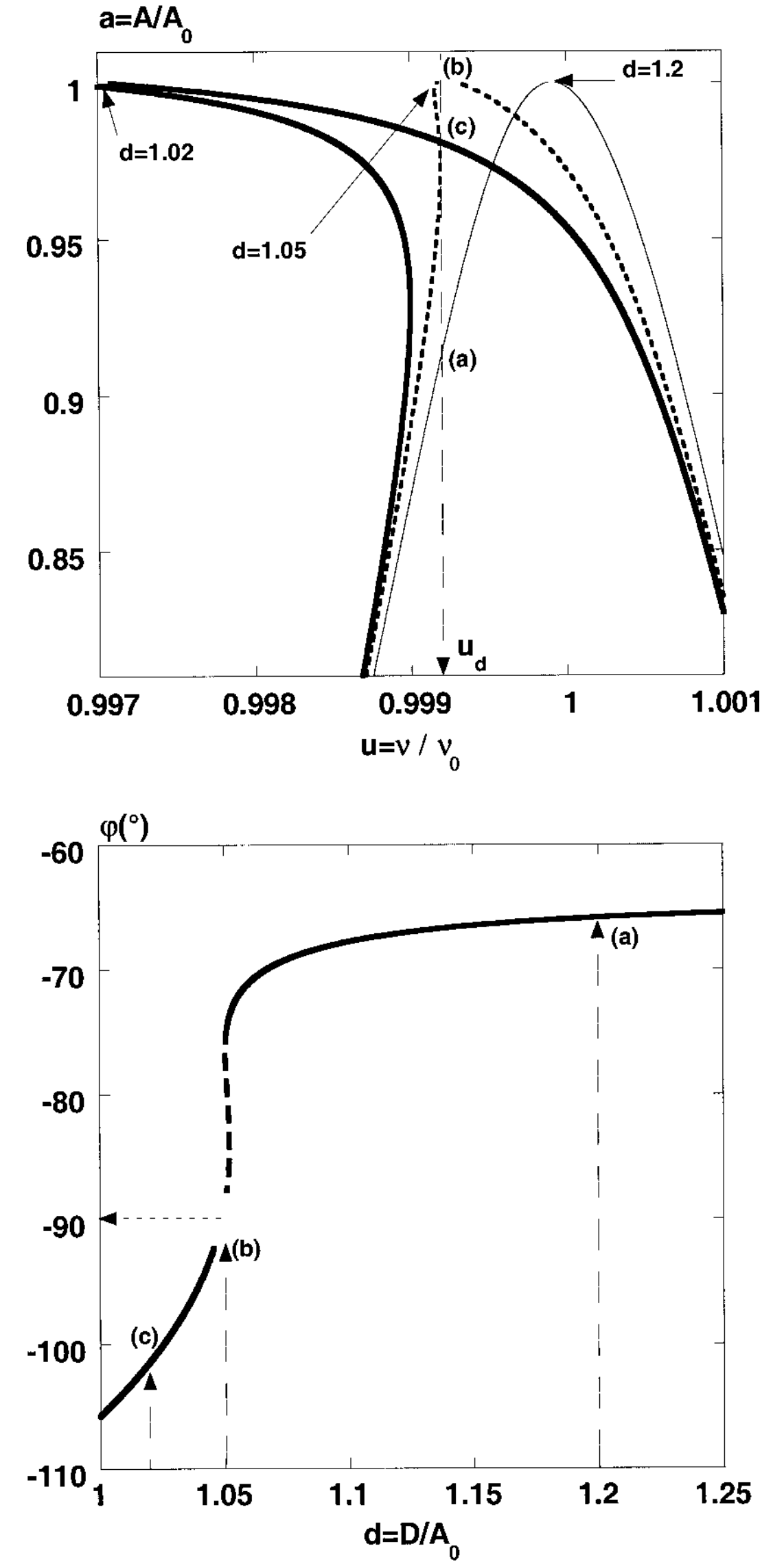

FIG. 3. (a) Distortion of the resonance peak under the action of the attractive field for three values of the distance $d$ [Eq. (8)]. The parameters are $A_0 = 15$ nm, $Q = 300$, and $\kappa_a = 1.48\times10^{-4}$. Letters indicate the transition between the stable behavior (domain (a), $d = 1.2$) with the free amplitude associated to $u_d = 0.9992$, $a_{\text{free}} = 0.91$, to the bistable behavior leading to an increase of the amplitude (domain (b), $d = 1.05$) and then the reduction of the amplitude (domain (c), $d = 1.02$). For clarity, only the upper part of the resonance peak is shown. (b) Analytical approach–retract curve of the phase illustrating the distortion of the resonance peak shown in Fig. 3(a). Arrows indicate the different points shown in Fig. 3(a) at $u_d = 0.9992$.

$$\cos(\varphi) = Qa(1-u^2) + \frac{4\sqrt{2}}{3\pi} Q\kappa_s a \left(1 - \frac{d}{a}\right)^{3/2}$$

$$\sin(\varphi) = -ua, \tag{9}$$

where $\kappa_s = k_s/k_c$ is the ratio between the contact stiffness and the cantilever stiffness. Solving Eq. (9) gives:

$$d_R = a\left\{1 - \left(\frac{3\pi}{4\sqrt{2}} \frac{Q(u^2-1) + \sqrt{\dfrac{1}{a^2} - u^2}}{Q\kappa_s}\right)^{2/3}\right\} \tag{10a}$$

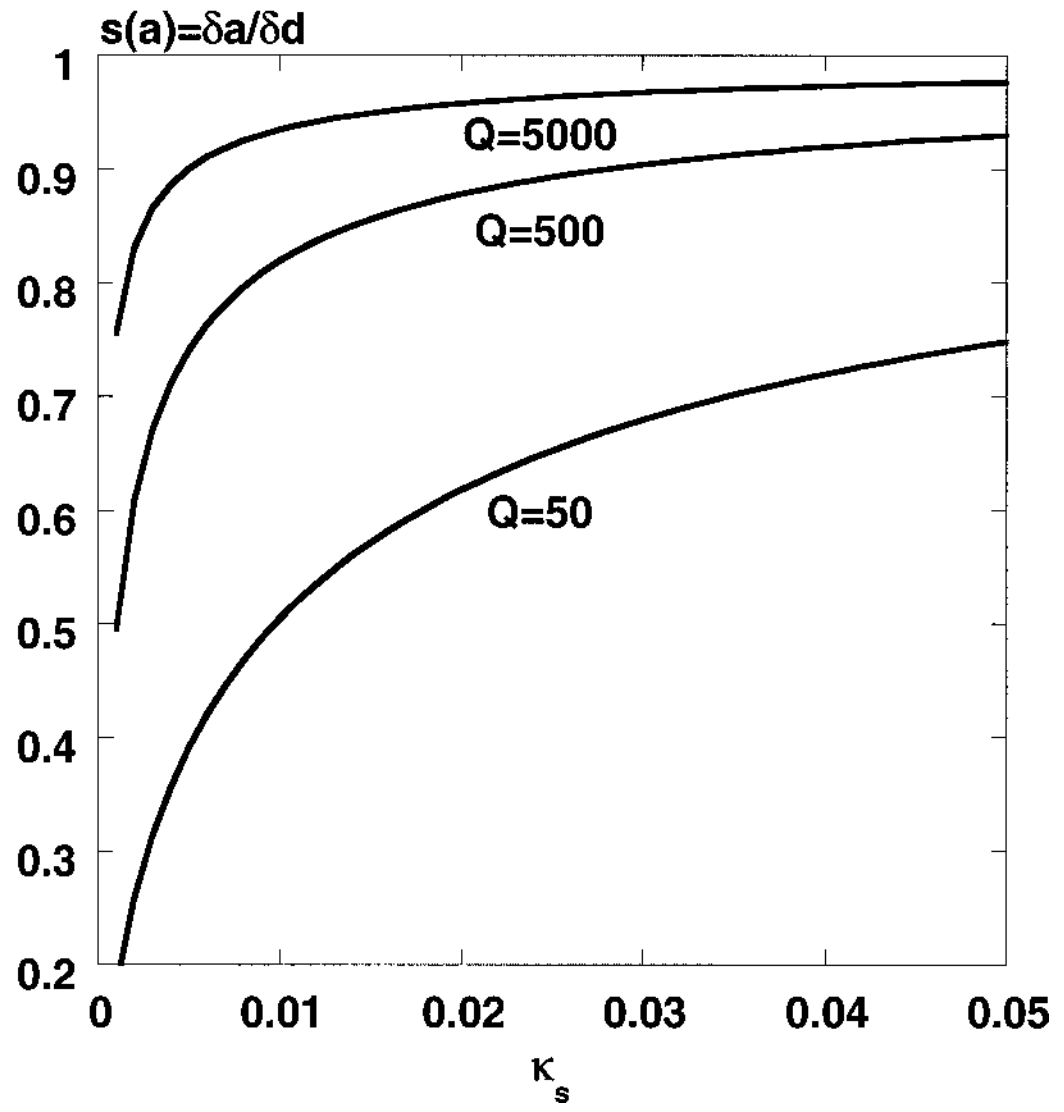

FIG. 4. Influence of the quality factor $Q$ on the slope in the IC case with uniquely a repulsive field [Eq. (11)]. The slopes as a function of the reduced stiffness $\kappa_s$ are calculated at the reduced amplitude $a = 0.65$. The other parameters are, $u = 0.9900$, $a_{\text{free}} = 0.707$, and $k_c = 10$ N m$^{-1}$.

$$\varphi_R = \arctan\left(\frac{u}{Q(u^2-1) - \dfrac{4\sqrt{2}}{3\pi} Q\kappa_s \left(1 - \dfrac{d_R}{a}\right)^{3/2}}\right). \tag{10b}$$

There is one physical branch of solutions in amplitude and in phase (see Appendix A). The key parameter is now $Q\kappa_s$ in place of $Q\kappa_a$.

The slope of the variation of the amplitude with $d, s(a) = \delta a/\delta d$, of an approach–retract curve is a fundamental parameter. Firstly, the slope contains information about the nanomechanical properties of the sample. Secondly, the slope at a given setpoint, that is at a given amplitude reduction at which an image is recorded, controls the vertical displacement of the piezoelectric actuator. Equation (10a) predicts a slope function of the $Q\kappa_s$ parameter. As a consequence, Eq. (10a) predicts that even for soft materials, a large $Q$ value could give a true topography of the surface, $s(a) = 1$, without any access to the mechanical properties of the sample. To be more quantitative, an analytical expression of the slope can be derived from Eq. (10a):

$$s(a) = \frac{1}{1 + \dfrac{\beta(a)}{(Q\kappa_s)^{2/3}}}, \tag{11}$$

where $\beta(a)$ is a complex expression of $a$, and $a_{\text{free}}$. The maximum value of $\beta(a)$ is around 5 and is reached for value of the amplitude close to $a_{\text{free}}$. Therefore for $Q\kappa_s \gg 10$, the slope is 1 over the whole domain of variation of the amplitude. Practically, one can evaluate the Young's modulus $G_s$ above which the slope is 1 whatever the value of the reduced amplitude $a$. In air, the quality factor is around 300. Choosing $k_c = 10$ N m$^{-1}$, we get $k_s \gg 0.33$ N m$^{-1}$ and with $\phi = 1$ nm, $G_s \gg 3.3\times10^8$ N m$^{-2}$.

Equation (11) predicts a nonlinear variation of the slope, but, within a selected range of amplitudes $a$ (see Sec. IV, Table I), a $k_s^{2/3}$ dependence is extracted that differs significantly from the one calculated for the static deflection mode. With the static mode, the slope, $s_{sd}$, measured with a force curve is given by:[27]

$$s_{sd}=\frac{1}{1+\dfrac{k_c}{k_s}}=\frac{1}{1+\dfrac{1}{\kappa_s}}, \tag{12}$$

besides a power law, the quality factor $Q$ plays a major role. To illustrate the influence of the quality factor, the variations of the slope as a function of $\kappa_s$ for $Q=50$, 500, and 5000 are reported in Fig. 4.

A comparison with a numerical simulation [the numerical simulation describing the intermittent contact is solved with the same procedure as the previous one, but the noncontact equation is replaced by a contact equation of the form, $\ddot{x}(t)+(\omega_0/Q)x(t)+\omega_0^2 x(t)=(f/m)\cos(\omega t)+(k_s/m)\times(x(t)-D)$] of the predicted variations of the amplitude and of the phase is given in Figs. 5(a) and 5(b). The numerical results validate the small indentation approximation. With $\kappa_s=0.0239$ and $Q=50$ a slope of about 0.22 is calculated whereas with $Q=5000$ and the same $\kappa_s$, the slope is 0.93.

The phase $\varphi_R$ is above $-90°$ during the entire approach–retract curve. When the oscillator touches the surface, the phase increases from its free value to 0°. This result emphasizes the difference between the NC state (phase below $-90°$) and the IC state (phase above $-90°$).

### B. Attractive and repulsive fields

A more realistic description of the IC includes an attractive interaction. To do so, we assume that the tip experiences the repulsive field during the two instants $[t_i;t_f]$, and an attractive van der Waals field during the time of the oscillation $[t_f;T=2\pi/\omega]$.

Unfortunately, because of the diverging behavior of the attractive field at $d=a$, it is difficult to obtain a simple, tractable, analytical solution. To get an analytical solution, we first assume that the attractive field acts during the whole period $T, t_f$ being a small part of the period. Then, the attractive contribution is evaluated for $d=a+\tilde{d}_c$, i.e., at the closest NC distance from the surface, where $\tilde{d}_c$ is the reduced coordinate of the contact distance $d_c=0.165\,\mathrm{nm}$:[23] $\tilde{d}_c=d_c/A_0$. With these conditions, we get:

$$\cos(\varphi)=Qa(1-u^2)+\frac{4\sqrt{2}}{3\pi}Q\kappa_s a\left(1-\frac{d}{a}\right)^{3/2}-\frac{Q\kappa_a}{6\sqrt{2}\tilde{d}_c^{3/2}\sqrt{a}},$$
$$\sin(\varphi)=-ua. \tag{13}$$

Solving Eq. (13) gives a cubic equation from which $d=f(a)$ is easily extracted.

$$aX^3-C=0\Leftrightarrow X=\left(\frac{C}{a}\right)^{1/3}, \tag{14}$$

with $C=(3\pi/4\sqrt{2})[Qa(u^2-1)+\sqrt{1-(ua)^2}+(Q\kappa_a/6\sqrt{2}\tilde{d}_c^{3/2})(1/\sqrt{a})]/Q\kappa_s$ and $X=\sqrt{1-d/a}$. Finally, the solutions are:

$$d_{AR}=a\left\{1-\left(\frac{C}{a}\right)^{2/3}\right\} \tag{15a}$$

$$\varphi_{AR}=\arctan\left(\frac{u}{Q(u^2-1)-\dfrac{4\sqrt{2}}{3\pi}Q\kappa_s\left(1-\dfrac{d_{AR}}{a}\right)^{3/2}+\dfrac{Q\kappa_a}{6\sqrt{2}(\tilde{d}_c a)^{3/2}}}\right). \tag{15b}$$

Here again, a unique physical branch of solution is given by Eqs. (15) (see Appendix A). As for the pure repulsive case, Eq. (15a) predicts a $(Q\kappa_s)^{2/3}$ dependence, but the relative importance of this parameter now becomes dependent of the magnitude of $(Q\kappa_s)$. While for small amplitudes (large $\kappa_a$) the slope is a complex function of $\kappa_a$ and $\kappa_s$, for large amplitudes (small $\kappa_a$), the OTCL becomes mostly sensitive to a pure repulsive field. Therefore, at large amplitudes, we expect the experimental curves to be suitably described with Eqs. (10).

A typical IC approach–retract curve requires to combine the solutions given by Eqs. (15) with the analytical results obtained with uniquely a pure attractive field [Eqs. (7a) and (7b)]. Theoretically, the equation of the surface is $d=a$, while a more realistic description preventing the diverging behavior uses $d=a+\tilde{d}_c$. A direct consequence is that branches of solution $d_{A+}(d_{A-})$ obtained with a pure attractive field might in fact touch the surface.

For small $\kappa_s$ (large $A_0$), at the bifurcation spot [point B, Figs. 6(a) and 6(b)] the branch of solution $d_{A-}(\varphi_{A-})$ is above the surface $d=a+\tilde{d}_c$, meaning that the tip jumps towards the upper IC branch $d_{AR}(\varphi_{AR})$. (point C). Then, the amplitude (phase) decreases (increases) (D). When the surface is retracted, the oscillator follows back the branch $d_{AR}(\varphi_{AR})$ until it reaches the maximum value of the amplitude (phase), which is solution of the equation $d_{AR}=d_{A+}$ $(\varphi_{AR}=\varphi_{A+})$ (point E) (or equivalently $d_{AR}=a+\tilde{d}_c$). At this point, as it is usually observed experimentally, the value of the amplitude is larger than the free one but smaller than thee resonance one for which $a = 1$. At the same time, the value of the phase is smaller than the free one but larger than the resonance one ($\varphi = -90°$). The structure of the whole curve exhibits a hysteresis cycle. At the end of the retract, the amplitude (phase) jumps back to its free value (F).

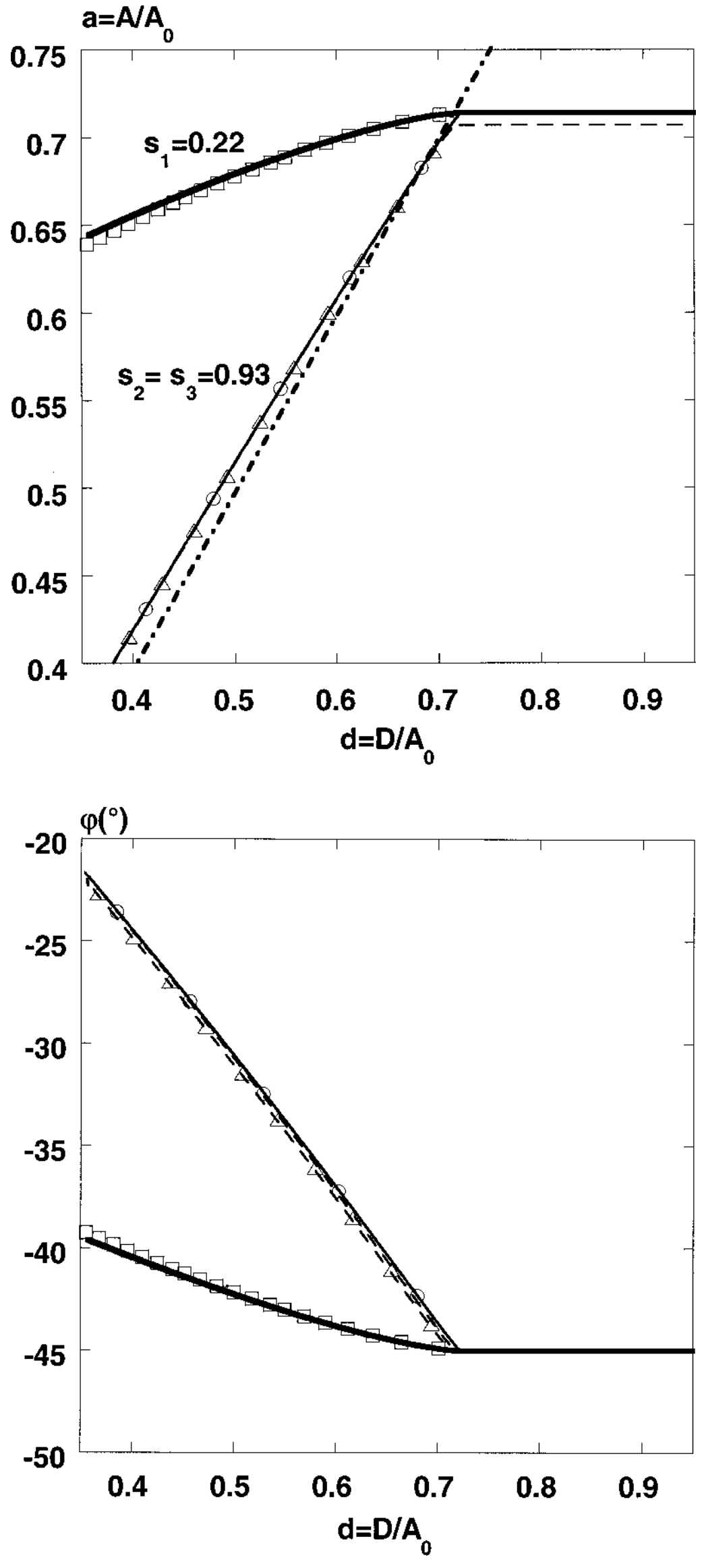

FIG. 5. (a) Evaluation of the small indentations approximation leading to Eqs. (10). The common parameters are $A_0 = 20\,\mathrm{nm}$, $k_c = 3.55\ \mathrm{N\ m^{-1}}$, $H = 0$ J, $R = 5$ nm. The location of the surface is given by the thick dashed-dotted line. Numerical results and curves calculated with Eq. (10a) have been performed with three different sets of parameters: (i) the analytical curve (thick continuous line) and the numerical simulation (empty squares) were computed with $Q = 50, u = 0.990\,05$, and $k_s = 0.085\ \mathrm{N\ m^{-1}}$ ($\kappa_s = 0.0239$ and $Q\kappa_s = 1.195$). The average slope extracted with a linear fit is $s_1 = 0.22$. Even for a small reduced stiffness $\kappa_s$, the comparison between the numerical results and the analytical one gives a good agreement down to the reduced amplitude $a = 0.4$. (ii) The analytical curve (thin continuous line) and the numerical simulation (empty circles) were obtained with $Q = 50, u = 0.990\,05$, and $k_s = 0.85\ \mathrm{N\ m^{-1}}$ ($\kappa_s = 0.239\ Q\kappa_s = 11.95$). (iii) The third set of parameters is $Q = 5000, u = 0.999\,90$, and $k_s = 0.085\ \mathrm{N\ m^{-1}}$ ($\kappa_s = 0.0239\ Q\kappa_s = 119.5$), analytical curve (thin dotted line) and numerical simulation (empty triangles). For these two sets of parameter, similar results are obtained on the whole range of variation of the reduced amplitude with a slope $s_2 = s_3 = 0.93$. (b) Comparison of the variations of the phase between the numerical simulations and the analytical curves calculated with Eq. (10b). The parameters are the same as the ones used for Fig. 5(a). The IC phases are above $-90°$.

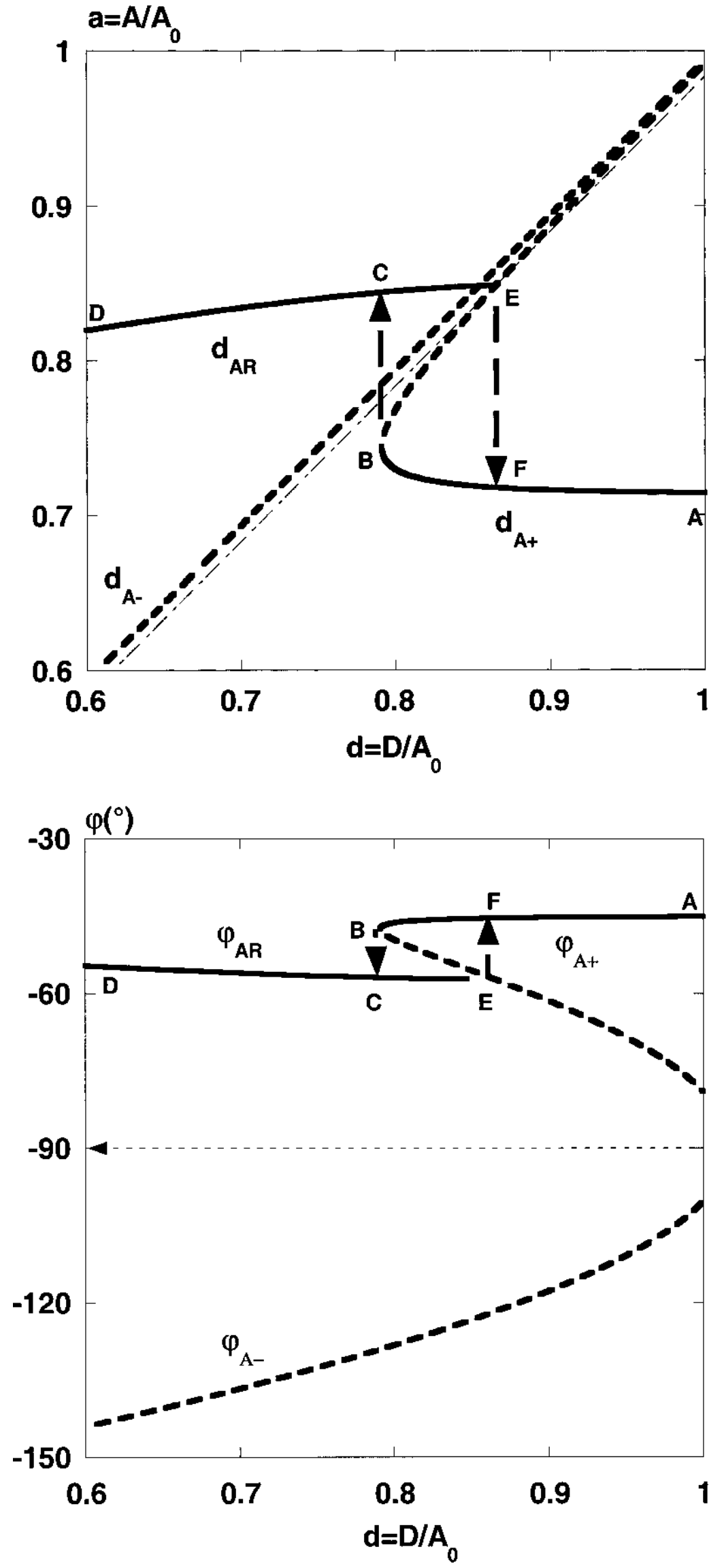

FIG. 6. (a) An analytical approach–retract curve showing the variation of the amplitude (thick continuous line). The attractive interaction is incorporated to describe the IC regime. The thick dashed lines give the unphysical branches. The branches $d_{A\pm}$ are calculated with Eq. (7a) with $A_0 = 10\,\mathrm{nm}$, $Q = 50, u = 0.9900$, and $\kappa_a = 10^{-4}$. The branch $d_{AR}$ is calculated with Eq. (15a) with the additional reduced stiffness $\kappa_s = 0.02$. The path BCEF describes the cycle of hysteresis. The thin dashed-dotted line gives the location of the surface. (b) Analytical approach–retract curve showing the variation of the phase associated to the variation of the amplitude shown in Fig. 6(a). $\varphi_{AR}$ is above $-90°$ over the whole approach and the whole retract of the surface.

At intermediate $\kappa_a$ values, more complex situations can be found that lead to a mixing of NC and IC situations. Figures 7(a) and 7(b) display most of the situations predicted. Since the bifurcation of the phase unambiguously discriminate between NC and IC stationary states, phase variations are uniquely reported.

For values of $\kappa_a$ not too large, the approach–retract curve presents the structure of Fig. 6(b) [Fig. 7(a): from $A_1$

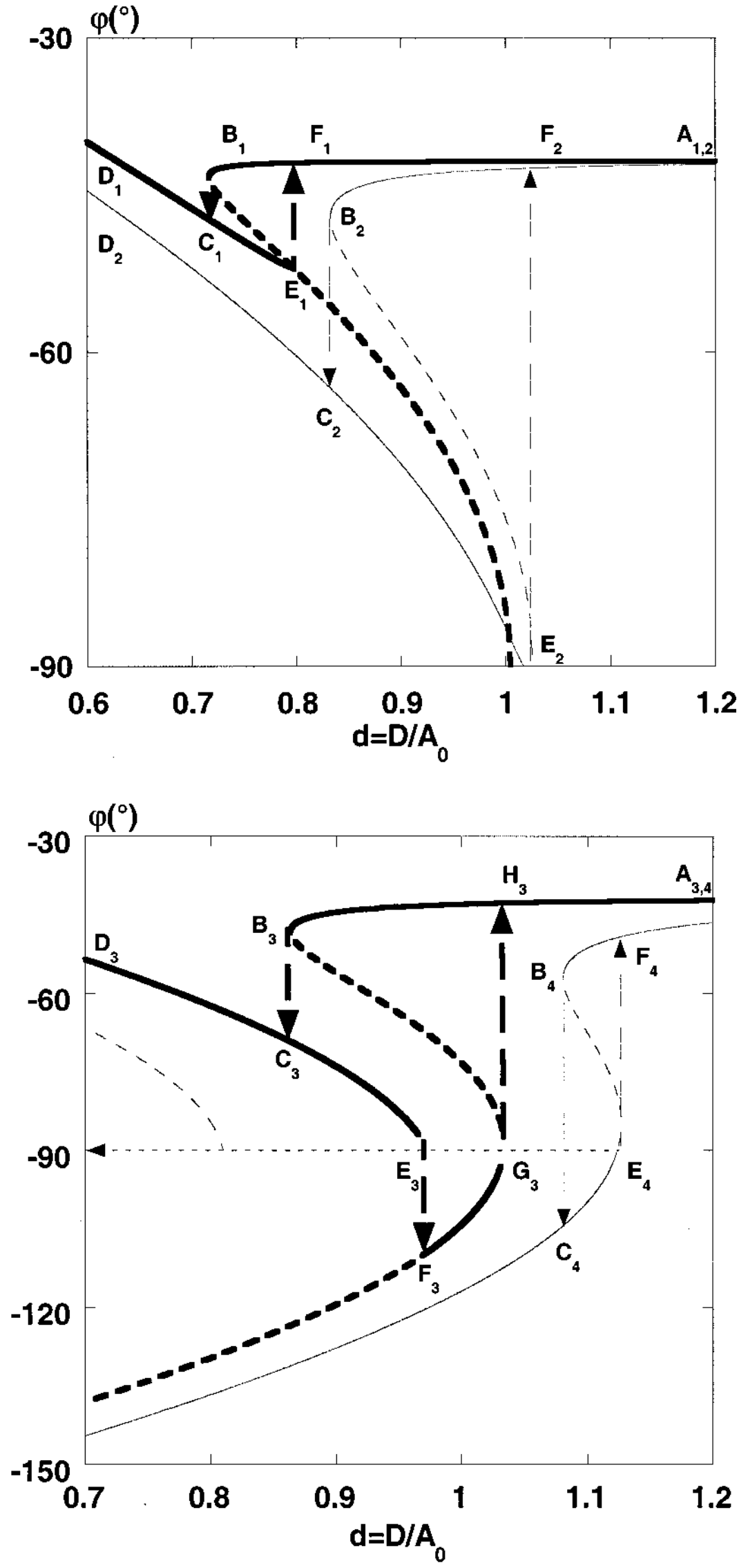

FIG. 7. (a) Analytical approach–retract curves showing the variations of the phase for two values of the drive amplitude [Eqs. (7b) and (15b)]. $A_0 = 20$ nm (thick lines) and $A_0 = 7$ nm (thin lines). Dashed lines represent the unstable branches and branches which cannot be reached. The parameters are $Q = 400$, $u = 0.9986$, $\kappa_s = 0.1$, and $\kappa_a = 3.5\times10^{-6}$, and $8.16\times10^{-5}$ for $A_0 = 20$ nm and $A_0 = 7$ nm, respectively. The path $B_iC_iE_iF_i$ describes the cycle of hysteresis. For this two-set of parameters, the branch $\varphi_A$ [Eq. (7b)] is never reached. (b) Same as in Fig. 7(a) but with two others values of $A_0$. $A_0 = 6$ nm (thick lines) illustrates a second unstable behavior (points $E_3$ to $F_3$) and $A_0 = 3$ nm (thin lines) illustrates a pure NC situation (branch $\varphi_A$) for which the IC is never reached. The parameters $\kappa_a$ are equal to $\kappa_a = 1.3\times10^{-4}$ and $1.04\times10^{-3}$ for $A_0 = 6$ nm and $A_0 = 3$ nm, respectively.

to $F_1$]. At a larger value of $\kappa_a$, the branches $\varphi_{A+}$ and $\varphi_{AR}$ join exactly at $\varphi = -90°$ [Fig. 7(a): from $A_2$ to $F_2$]. This is still a situation of IC.

When $\kappa_a$ is further increased, again during the approach the OTCL jumps down to the $\varphi_{AR}$ branch of solutions [Fig. 7(b), points $A_3$ to $C_3$], while during the retract $\varphi_{AR}$ doesn't join $\varphi_{A+}$ anymore. During the retract, the phase associated to the $\varphi_{AR}$ branch reaches the value of the resonance one: $-90°$ ($E_3$), then jumps down to the $\varphi_{A-}$ branch ($F_3$) before reaching the $\varphi_{A+}$ branch ($H_3$). This situation gives a second

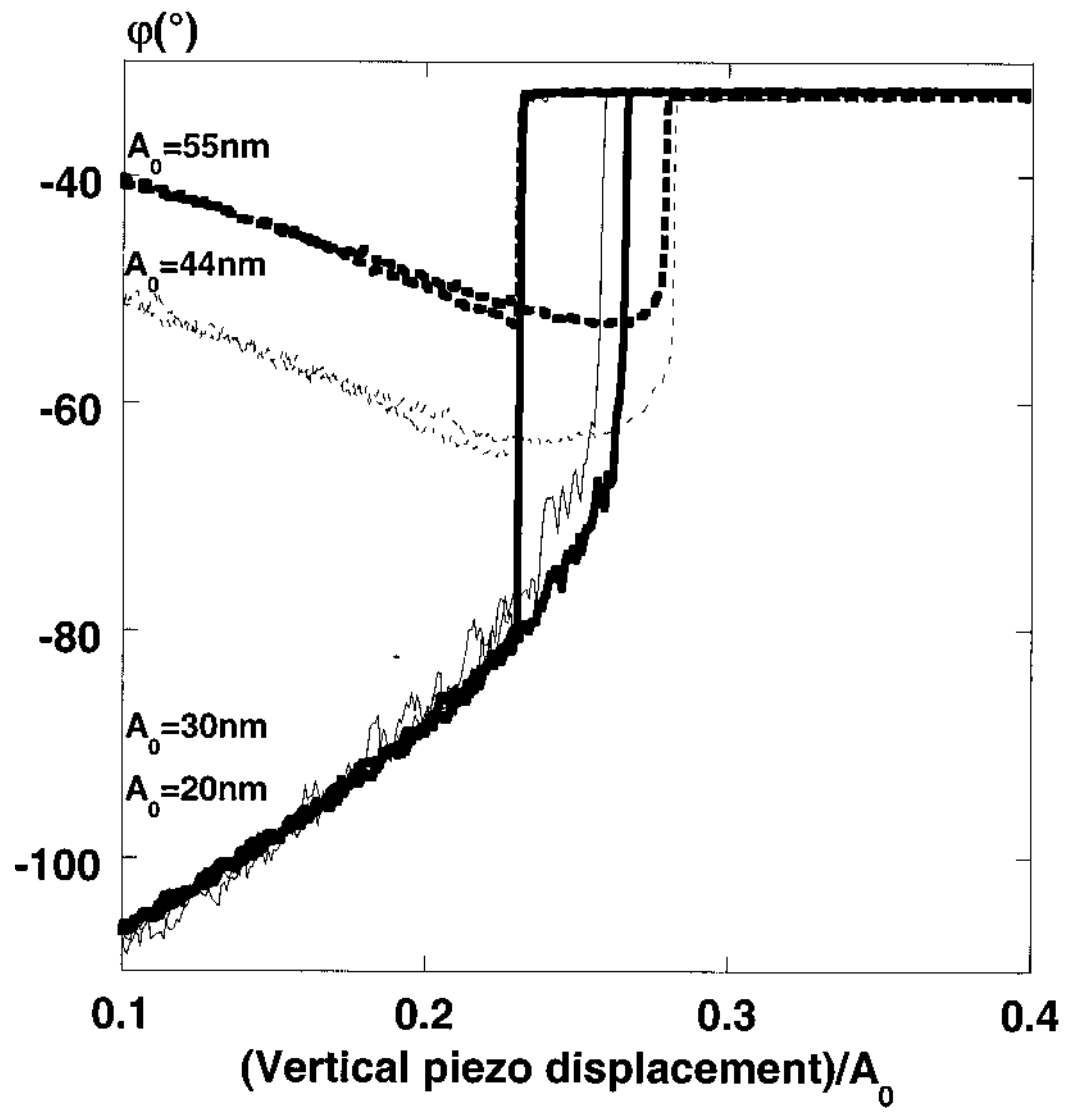

FIG. 8. Experimental approach–retract curves showing the variation of the phase on a polystyrene polymer film of molecular weight $M_w = 200\,000$. The experimental drive amplitudes give $A_0 = 55$, 44, 30, and 20 nm. The OTCL used is the OTCL1 (Appendix B). The zero of the vertical location of the surface is set arbitrarily (see Appendix B). The NC stationary state occurs at a value as high as $A_{\text{free}} = 0.707\times A_0$, with $A_0 = 30$ nm, thus suggests that the tip's radius is large (see text)

bistable behavior of the oscillator with an intermediate NC situation.

For large values of $\kappa_a$, the branch of solutions $\varphi_{AR}$ is too far to be reached during the approach [Fig. 7(b), point $B_4$]. The stable branch is uniquely the NC one and the OTCL follows the usual NC path [Fig. 7(b), points $C_4$ to $A_4$].

## IV. DISCUSSION

Because of the nonlinear behavior of the OTCL and of the mixing of IC and NC stationary states, complex situations can happen as those shown in Fig. 7(b). But, while some of them will not be easy to observe experimentally, others, theoretically forbidden, will occur because of tip (or surface) pollution. In the present part are given some typical results, showing that the observed variations of the phase allow the IC and NC situations to be easily discriminated. Also some unpredicted situations are presented. Finally, we briefly discuss two points that have not been analytically solved. One is the case of dissipating samples, a rather important point for soft materials, the second concerns numerical results in which are included the Hertz model to describe the IC state.

The experimental results shown in Figs. 8 and 9 were performed in a glove box on silica surfaces and polystyrene polymer films of molecular weight $M_W = 200\,000$. The silica surfaces were prepared as described in Ref. 28. Both silica and polymer surfaces can be considered as hard surfaces, the polymer film for high molecular weights being mechanically inert under the action of the oscillating tip in the attractive regime.[16] In the glove box, the p.p.m. of water is achieved so that capillary forces are negligible (see also Appendix B).

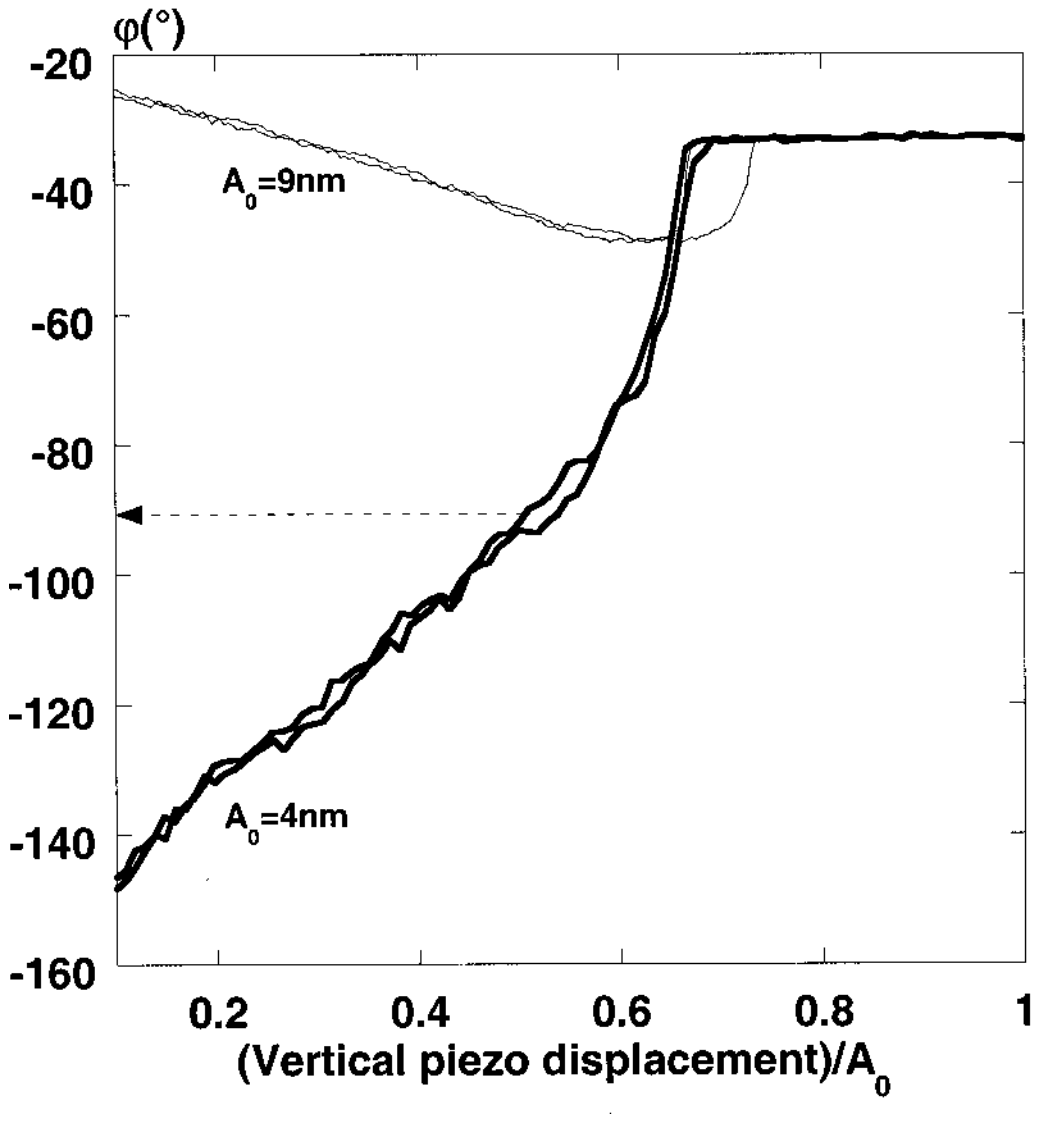

FIG. 9. Experimental approach–retract curves performed with the OTCL2 on a silica surface. The experimental conditions are given in Appendix B. At $A_0 = 9$ nm, $A_{\mathrm{free}} = 6.4$ nm, the OTCL remains in a well-defined IC stationary state. The NC state is observed for values equal or below $A_0 = 4$ nm ($A_{\mathrm{free}} = 2.8$ nm). For such a low value of the amplitude, the instability has disappeared and the variation of the phase resembles the theoretical one given in Fig. 2(b).

The general trend predicted by Eqs. (15) is in good agreement with the one measured. The relative importance of the dimensionless parameter $Q\kappa_a$ can be checked by varying the drive amplitude,[19] thus $A_0$, or using a large tip radius (or a crashed tip), a larger HR product. With a crashed tip, a pure attractive regime shown by the variation of the phase occurs at a $A_0$ value as high as 30 nm (Fig. 8). For NC stationary states exhibiting a bifurcation, as soon as the OTCL had crossed the bifurcation point, the phase varies slowly and its variation is practically independent of the $A_0$ value, thus a similar variation is observed with $A_0 = 20$ nm.

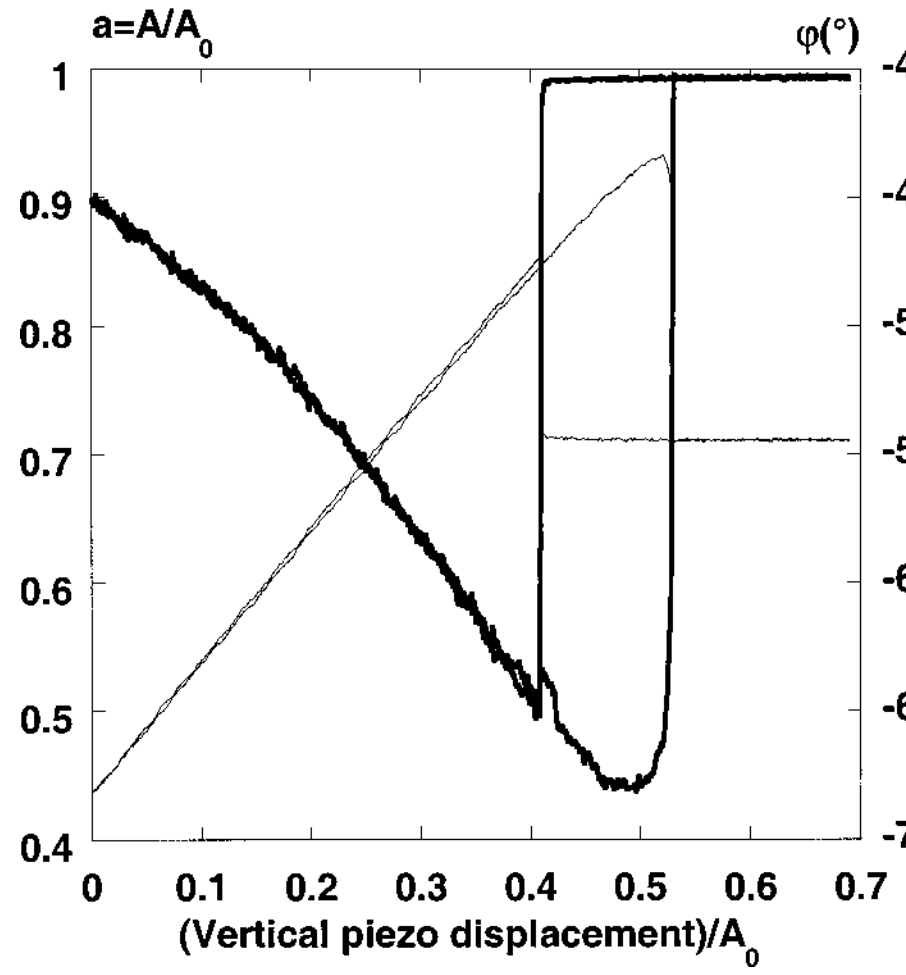

FIG. 10. Variation of the amplitude and the phase observed on a mica surface with the OTCL3 with $A_0 = 41$ nm. Phase and amplitude indicate an IC state.

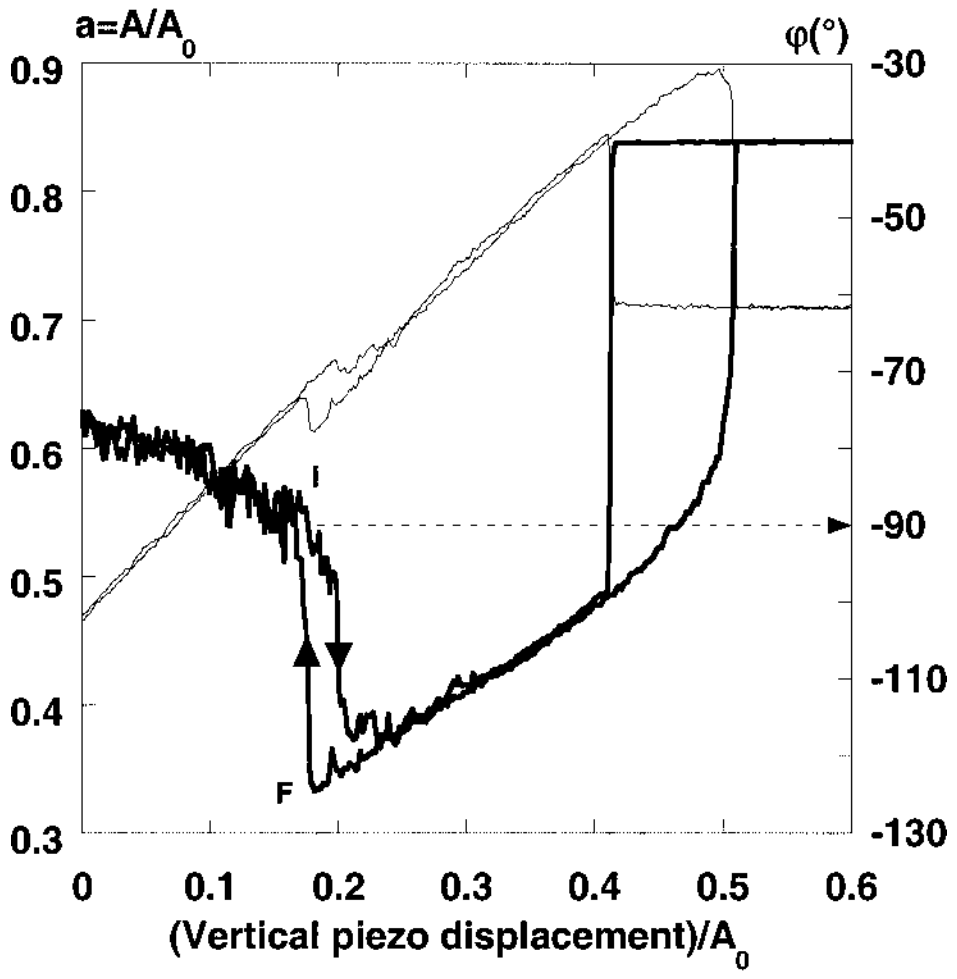

FIG. 11. Variation of the amplitude and the phase observed on a mica surface with the OTCL3 with $A_0 = 32$ nm. The NC branch is the stable one [see the fourth case in Fig. 7(b)], but an instability occurs (point F) and the IC branch is reached (point I). Such a jump is not predicted. Here is involved a possible pollution of the tip and/or of the surface (see text).

With a second OTCL, the pure attractive regime uniquely occurs for small values of $A_0$ (below $A_0 = 4$ nm,) from which is deduced that the tip is much smaller (Fig. 9). For such a small tip radius, a bifurcation structure is more difficult to observe and the phase crosses almost continuously the $-90°$ value [as shown theoretically, Fig. 2(b)].

The situations showing a jump from an IC branch to a NC one during the retraction of the surface [path $E_3F_3G_3H_3$, Fig. 7(b)] were not experimentally encountered. The reason is probably that this type of jump occurs at a spot close to the one corresponding to the usual instability jumping back to the free values of the phase and amplitude. Nevertheless, bifurcation structures resembling those can be found for soft materials. In that case, the bifurcation is due to the growth of a viscoelastic nanoprotuberance under the action of the attractive field of the OTCL.[16]

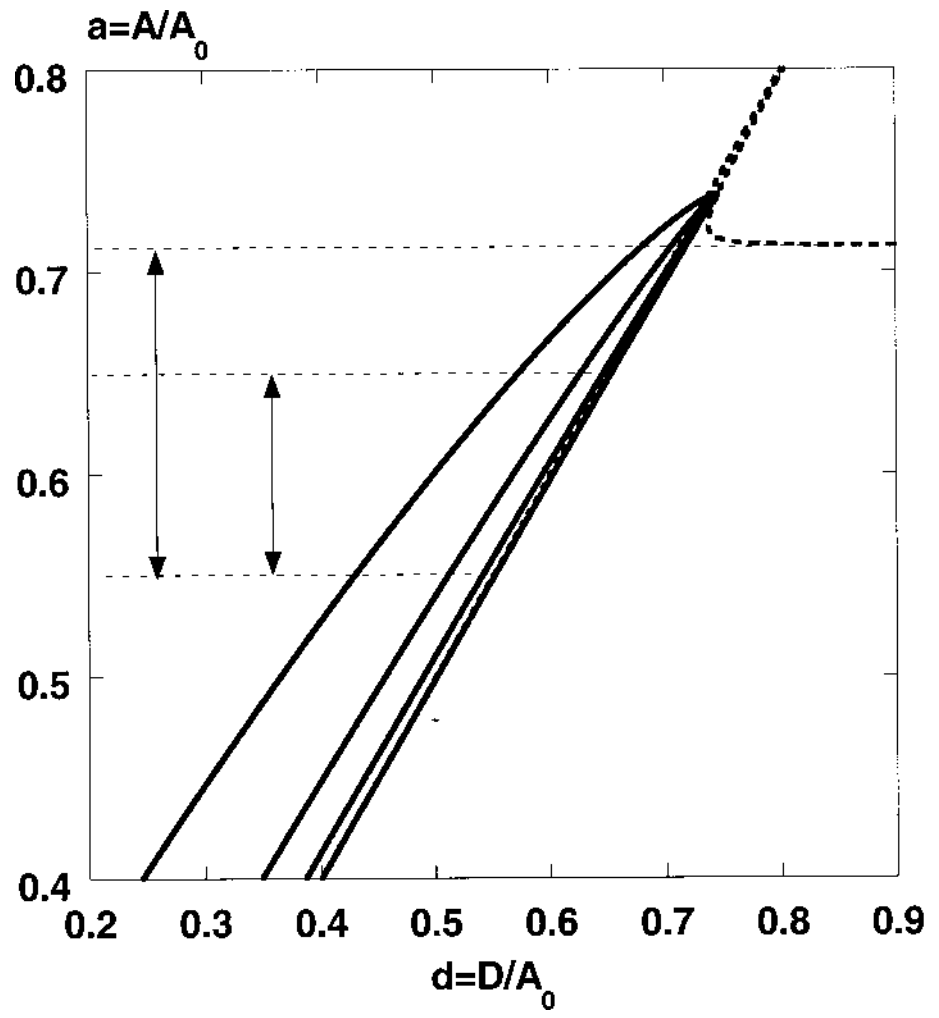

FIG. 12. Variation of the amplitude as a function of the OTCL-surface distance calculated with Eq. (15a). The parameters are the ones used in the Table. The dotted curve corresponds to the pure attractive case [Eq. (7a)]. The continuous lines from the right to the left have been calculated with the Young Modulus 100, 10, 1, and 0.1 GPa, respectively. The arrows indicate the domain of variation used to fit the curve with a linear function (see Table).

TABLE I. Comparison between the results obtained with two numerical simulations and the curve calculated with Eq. (15a). Variations of the IC slope as a function of $\kappa_s$. Two domains of variations of the reduced amplitude $a$ are chosen to be fitted with a linear approximation (from the free amplitude 0.707 to 0.55 and from 0.65 to 0.55). The function used to extract the exponent $\alpha$ is $s=1/(1+C/\kappa_s^{\alpha})$.

| | Hertz model | | $k_s x$ model | | Equation (15a) | |
|---|---|---|---|---|---|---|
| $\kappa_s$ | 0.707→0.55 | 0.65→0.55 | 0.707→0.55 | 0.65→0.55 | 0.707→0.55 | 0.65→0.55 |
| 18.99 | 0.99 | 0.99 | 0.98 | 0.98 | 0.975 | 0.98 |
| 4.08 | 0.95 | 0.96 | 0.935 | 0.95 | 0.93 | 0.95 |
| 0.87 | 0.87 | 0.89 | 0.83 | 0.86 | 0.84 | 0.86 |
| 0.19 | 0.69 | 0.73 | 0.65 | 0.69 | 0.65 | 0.70 |
| Exponent $\alpha$ | 0.81 | 0.77 | 0.71 | 0.68 | 0.67 | 0.66 |

It's far beyond the scope of the present work to discuss in detail such a behavior. Moreover, the study of polymer films made of tribloc that exhibit a regular structure at the mesoscopic scale with different mechanical responses[10,11,12] is more suitable. As a matter of fact such a material, and in general the mixing of hard and soft materials, exhibit a wide variety of situations that requires a specific analysis.[14,15]

The latter example concerns the study of a mica surface in the glove box with a rather large tip (Figs. 10 and 11). The structure observed is reminiscent of the fourth case given in Fig. 7(b) (path $B_4C_4E_4F_4$). At a very large drive amplitude, $A_0=41$ nm, the usual variations of the IC state are observed (Fig. 10). At a slightly lower one, $A_0=32$ nm, the OTCL follows the NC branch then suddenly jumps to the IC branch to reach the value $\varphi=-90°$. During the retract, the OTCL follows the IC branch until $\varphi=-90°$ is reached, then jumps down to the NC branch (Fig. 11). This situation corresponds to the fourth case for which the NC branch is the stable one and the IC state should normally not be reached [Fig. 7(b)]. In other words, experimental results similar to the ones shown in Fig. 8 ($A_0=30$ and 20 nm), where uniquely the NC branch is available, should be observed.

To explain the observed IC state, the most reasonable hypothesis is the occurrence of pollution either on the tip or on the surface that allows an IC situation to happen quenching the OTCL in the IC state. This can be explained with the fact that the average attractive field between the OTCL and the surface varies as $1/\sqrt{A}$.[16] Therefore when the amplitude decreases, the attractive field increases so that materials can be pumped, and creates IC situations.

Not included in the above analysis is the influence of an additional dissipation when an IC state takes place. As stated above, great care must be taken to discuss phase variation as the result of a dissipating process. In any case, the introduction of an additional dissipation will make almost impossible the search for an analytical solution without the use of assumptions making the whole attempt questionable. Numerical simulations can help to identify the role of dissipation. Qualitatively some information can be used[12,13,14] but even in that case, the phenomenological approach does not give a clear meaning of what is really occurring. The additional dissipation can be issue of the bulk viscoelastic properties of the sample, requiring that the volume involved is at least of the cubic micrometer.[29] Such a huge volume is certainly not involved in IC situations. If the dissipate process occurs at the interface between the tip and the surface, there is no simple analytical solutions and the description of the dissipating process remains quite complicated even in the case of the static deflection mode.[27]

Finally, a question arises about the validity of the simple harmonic model employed to describe the repulsive field. A somewhat more complicated approach is to introduce the Hertz model, which gives a nonlinear dependence between the indentation depth and the elastic force.[25] The Hertz model describes an elastic contact between the tip and the sample and provides an evaluation of the contact stiffness as a function of the indentation depth. The introduction of the Hertz model doesn't lead to a simple analytical solution.[14] The results presented above are preserved, structure of the bifurcation, discrimination between the IC and NC states, and the main difference concerns the slope dependence as a function of the contact stiffness. The exponents found are slightly above the ones obtained with the harmonic approximation (Table I and Fig. 12).

Note that for reasonable values of $\kappa_a$, numerical and analytical results give a power law close to 2/3. A simple analysis of Eq. (15a) indicates that with large amplitudes, the influence of the attractive field is reduced as $\kappa_a$ goes asymptotically to zero, thus allowing the slope to have a simplest form as the one given in Eq. (11). Therefore, with a high amplitude it becomes easier to access the local mechanical properties more quantitatively. The other point is the important role of the quality factor $Q$. As shown with Eq. (11) and Eq. (15a), the higher the quality factor, the larger the product $Q\kappa_s$. Therefore to be very sensitive with an OTCL to access at the nanomechanical properties of the sample requires a low $Q$ factor, while to access the topographic structure requires a high $Q$ factor and a high amplitude. The unique way to check the validity of these equations is to use a vacuum chamber in which the magnitude of the quality factor can be varied.

## V. CONCLUSION

The aim of the present work was to extract analytical expressions describing the nonlinear behavior of an OTCL in the tapping mode. Three situations are considered: the pure attractive interaction, the pure repulsive interaction, and a mixing of the two. The analytical solutions give the variation of the amplitude and of the phase as a function of the OTCL-surface distance. The general evolutions predicted are in good agreement with the observed ones. The NC and IC stationary states can be discriminated by recording the variation of the phase. The bifurcation from a monostable to a bistable state is identified in the pure attractive regime. In the intermediate regime, because of the nonlinear behavior of the OTCL and the mixing of IC and NC stationary states, the structures of the bifurcation are more complex. Experimentally, it is easy to discriminate between the different situations by varying the magnitude of the drive amplitude. The contribution of the attractive force can be significantly reduced through the increase of the drive amplitude leading to an almost pure repulsive case, while a NC stationary state becomes dominant at small drive amplitudes. The quality factor $Q$ is shown to be a key parameter. For large values of $Q$ and large amplitudes a true topography is expected, while with a smaller value of $Q$, the OTCL becomes more sensitive to the local mechanical properties of the surface. Therefore, it does appear interesting to perform experiments on soft materials in a vacuum chamber in order to vary the quality factor.

## APPENDIX A: THE VARIATIONAL PRINCIPLE

### 1. Pure attractive field

The principle of least action specifies that the action $S[x(t)]$ being a functional of the path $x(t)$ between two instants is extremal.

$$\delta\{S[x(t)]\}=\delta\left\{\int_{t_i}^{t_r} L(x;x;t)dt\right\}=0, \tag{A1}$$

where $L$ is the Lagrangian of the system. The aim of the use of a variational principle is to employ a trial function $x(t)$ allowing us to perform an analytical treatment. According to the fact that the oscillator essentially exhibits a harmonic behavior, the trial function used has the harmonic stationary form of Eq. (2), and therefore the action is defined on one oscillating period. The Lagrangian of the OTCL is defined as:

$$\begin{aligned} L&=T-U+W\\ &=\frac{1}{2}m\dot{x}^2(t)-\left[\frac{1}{2}m\omega_0^2x^2(t)-x(t)f\cos(\omega t)\right.\\ &\quad\left.-\frac{HR}{6(D-x(t))}\right]-\frac{m\omega_0}{Q}x(t)\underline{\dot{x}(t)}, \end{aligned} \tag{A2}$$

with:

$$x(t)=A\cos[\omega t+\varphi]. \tag{A3}$$

The parameters of the path are $A$ and $\varphi$, and the principle of least action $\delta S=0$ gives a set of two partial differential equations:

$$\frac{\delta S}{\delta A}=0, \quad \frac{\delta S}{\delta\varphi}=0. \tag{A4}$$

Three terms are dissociated for the calculations of the action:

$$S=S_0+S_{\mathrm{int}}+S_{\mathrm{dissip}}, \tag{A5}$$

where $S_0$, $S_{\mathrm{int}}$, and $S_{\mathrm{dissip}}$ are, respectively, the action of the harmonic oscillator without dissipation, the action of the interacting term, and the action of the dissipating term.

$$S_0=\int_0^{2\pi/\omega}\left[\frac{1}{2}m\dot{x}(t)^2-\frac{1}{2}m\omega_0^2x(t)^2+x(t)f\cos(\omega t)\right]dt \tag{A6a}$$

$$S_{\mathrm{int}}=\int_0^{2\pi/\omega}\left[\frac{HR}{6(D-x(t))}\right]dt \tag{A6b}$$

$$S_{\mathrm{dissip}}=\int_0^{2\pi/\omega}\left[-\frac{m\omega_0}{Q}x(t)\underline{\dot{x}(t)}\right]dt. \tag{A6c}$$

The calculation of $S_0$ gives:

$$S_0=\frac{\pi m}{2\omega}A^2(\omega^2-\omega_0^2)+\frac{\pi f}{\omega}A\cos(\varphi). \tag{A7}$$

$\dot{x}(t)=-\underline{A}\omega\sin(\omega t+\underline{\varphi})$ is calculated along the physical path,[21] with $\underline{A}$ and $\underline{\varphi}$ being fixed for the minimization. The integration gives:

$$S_{\mathrm{dissip}}=-\frac{\pi m\omega_0}{Q}A\underline{A}\sin(\varphi-\underline{\varphi}). \tag{A8}$$

The calculation of $S_{\mathrm{int}}$ requires a more tedious, but straightforward, calculation:

$$S_{\mathrm{int}}=\frac{\pi HR}{3\omega D\sqrt{1-e^2}}, \tag{A9}$$

with $e=A/D=a/d<1$. Adding the expressions (A7), (A8), and (A9) gives the complete action:

$$\begin{aligned} S=&\frac{\pi m}{2\omega}A^2(\omega^2-\omega_0^2)+\frac{\pi f}{\omega}A\cos(\varphi)+\frac{\pi HR}{3\omega\sqrt{D^2-A^2}}\\ &-\frac{\pi m\omega_0}{Q}A\underline{A}\sin(\varphi-\underline{\varphi}). \end{aligned} \tag{A10}$$

The minimization with respect to $A$ gives:

$$\begin{aligned} \frac{\partial S}{\partial A}=&\frac{\pi m}{\omega}A(\omega^2-\omega_0^2)+\frac{\pi f}{\omega}\cos(\varphi)+\frac{\pi HRA}{3\omega(D^2-A^2)^{3/2}}\\ &-\frac{\pi m\omega_0}{Q}\underline{A}\sin(\varphi-\underline{\varphi}), \end{aligned} \tag{A11}$$

whereas the minimization with respect to $\varphi$ gives:

$$\frac{\partial S}{\partial\varphi}=-\frac{\pi f}{\omega}\underline{A}\sin(\varphi)-\frac{\pi m\omega_0}{Q}A\underline{A}\cos(\varphi-\underline{\varphi}). \tag{A12}$$

Since the underlined variables are calculated along the physical path, $A=\underline{A}$ and $\varphi=\underline{\varphi}$. With these conditions, solving $\partial S/\partial A=0$ and $\partial S/\partial\varphi=0$ and with the external drive force written as $f=m\omega_0^2A_0/Q$, (A11) and (A12) finally give the two equations of motion:

$$\cos(\varphi)=Qa(1-u^2)-\frac{aQ\kappa_a}{3(d^2-a^2)^{3/2}}$$
$$\sin(\varphi)=-ua. \tag{A13}$$

To get Eq. (7), the system is solved using the trigonometric relationship: $\cos(\varphi)=\pm\sqrt{1-\sin^2(\varphi)}$. Those two branches are defined for every set of parameters and therefore both have to be conserved.

### 2. Pure repulsive field

The interacting term is given by:

$$S_{\text{int}}=2\times\frac{k_s}{2}\int_0^{\arccos(D/A)/\omega}(x(t)-D)^2\,dt. \tag{A14}$$

Calculating the action, then using a Taylor's expansion of arccos(*D*/*A*) for $A\approx D$, the calculations finally lead to Eq. (9):

$$\cos(\varphi)=Qa(1-u^2)+\frac{4\sqrt{2}}{3\pi}Q\kappa_s a\left(1-\frac{d}{a}\right)^{3/2}$$
$$\sin(\varphi)=-ua. \tag{A15}$$

With $\cos(\varphi)=\pm\sqrt{1-\sin^2(\varphi)}$, expression (A15) also gives two branches of variations. The fractional power law 2/3 in Eq. (10a) implies to choose $\cos(\varphi)=+\sqrt{1-\sin^2(\varphi)}$, thus leading to only one physical branch.

### 3. Both attractive and repulsive fields

The interacting term is given by:

$$S_{\text{int}}=2\times\left\{-\frac{HR}{6}\int_{[\arccos(D/A)]/\omega}^{\pi/\omega}\frac{dt}{(D-x(t))}+\frac{k_s}{2}\int_0^{[\arccos(D/A)]/\omega}(x(t)-D)^2\,dt\right\}. \tag{A16}$$

The attractive NC part is evaluated using the expression (A9). Then, a Taylor's expansion of arccos(*D*/*A*) for $A\approx D$ is used, and a physical surface given by $d=a+\tilde{d}_c$ is introduced (see text). The calculations finally lead to the couple of equations:

$$\cos(\varphi)=Qa(1-u^2)+\frac{4\sqrt{2}}{3\pi}Q\kappa_s a\left(1-\frac{d}{a}\right)^{3/2}-\frac{Q\kappa_a}{6\sqrt{2}\tilde{d}_c^{3/2}\sqrt{a}}$$
$$\sin(\varphi)=-ua. \tag{A17}$$

Here again, because of the fractional power law, only one physical branch of solution is available.

## APPENDIX B: EXPERIMENTAL CONDITIONS

The experimental approach–retract curves were measured with the tapping mode of a Nanoscope III from Digital Instruments (Santa Barbara, CA). The cantilevers (Nanoprobe) have a typical resonance frequency around 300 kHz and a quality factor between 250 and 550 (see Table B1).

Two piezoactuators are needed to perform experiments with the tapping mode: one allowing the microlever to vibrate at a given drive frequency with a given drive amplitude, and a second piezo to move the sample. During an approach–retract curve, the second piezo displaces the sample along the *Z*-axis while the other keeps constant the drive frequency and the drive amplitude.

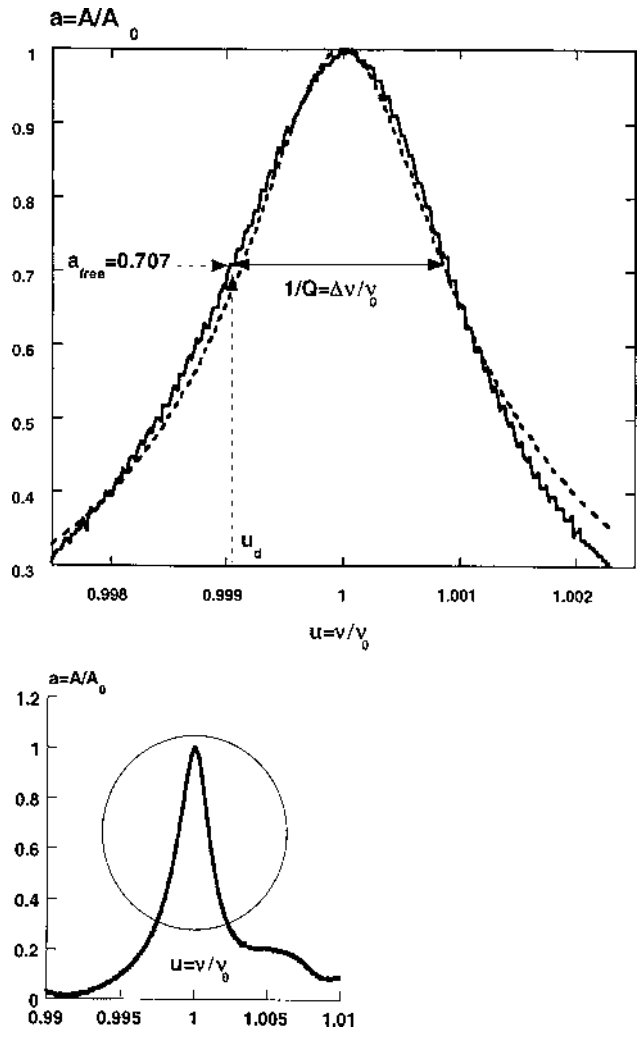

FIG. B1. Resonance curve with fit corresponding to Eq. (4a).

### 1. Technical remarks: preparation of the experiment

The resonance frequency and the quality factor are extracted from measurements of the resonance peak (Cantilever Tune mode on the NIII). Recording of the resonance peaks are made at distances from the surface between 500 nm and 1 $\mu$m. For distances above 10 $\mu$m, it has been shown that parameters of the OTCL vary.[12,30] Change of the resonance amplitude and frequency had been attributed to the decrease of the thickness of a layer of air between the OTCL and the surface. The OTCL properties become stable as soon as the distance between the OTCL and the surface is around or smaller than one micrometer.

The quality factor $Q$ is measured at $A=A_0/\sqrt{2}$ (at $a=1/\sqrt{2}=0.707$ with the reduced coordinates). For large values of $Q$, the quality factor is given by $Q=\nu_0/\Delta\nu$, with $\Delta\nu$ the frequency width at this amplitude. Using the reduced coordinates, a resonance curve is presented in Fig. B1 with the corresponding fit [Eq. (4a)]. In many cases the resonance curves may exhibit an asymmetric shape and several bumps in the curve tails. Practically, the undesirable bumps can be resumed by clearing the lodgment of the cantilever and moving the cantilever inside its lodgment.

At the very beginning, the resonance peak is often asymmetric. Typically, after three hours the asymmetry has almost completely disappeared and the resonance peak recovers a Lorentzian shape. As such a kinetic effect was observed, it might be due to a thermal equilibration of the OTCL under the action of the laser beam focalized at its upper extremity.

TABLE B1. Experimental parameters and samples investigated. PS film is a polystyrene polymer film of molecular weight $M_W = 200\,000$.

| OTCL | Resonance frequency (kHz) | Drive frequency (kHz) | Quality factor $Q$ | Sample |
|---|---|---|---|---|
| OTCL1 | 300.069 | 299.718 | 400 | PS film |
| OTCL2 | 278.809 | 278.490 | 450 | Silica surface |
| OTCL3 | 234.987 | 234.663 | 350 | Mica surface |

After a few hours, the resonance curve becomes stable and the OTCL properties do not evolve. The OTCL for which it is impossible to extract a harmonic behavior (several oscillating modes, large asymmetry remaining) are rejected.

Once the OTCL had reached the thermal equilibrium with the laser beam, the stability of the whole system and the reproducibility of the experiments were excellent. Only drifts of the resonance frequency of around a few tens hertz were observed over a period of several weeks. This is particularly true in the glove box, while in-air variations of the resonance frequency and quality factor were observed over several days, those variations being probably due to change of the ambient conditions.

A drift of a few tens hertz does not alter the approach–retract curves. Several identical measurements were done time-to-time showing an identical variation of the amplitude and the phase as a function of the OTCL surface distance. When it occurs, the main change is due to a variation of the tip apex when very soft materials are investigated, meaning that some polymer had coated the tip.

## 2. Data treatment

At a given drive amplitude the used amplitude is chosen with the cantilever tune such that $A_{\text{free}} = 0.707 A_0$. The corresponding phase data were typically between $-30°$ and $-40°$ (To make the phase varying between 0° and $-180°$, subtraction of the recorded data of $-90°$ was done). Dispersion of the values of the phase is partly due to the accuracy to set the magnitude of $A_{\text{free}}$, and the way the phase offset is fixed with the cantilever tune.

The calibration of the amplitude is achieved by making series of approach–retract curves with different drive amplitudes on a hard surface chosen as a reference (usually a silica surface). The hard surface makes the slope of the IC equal to 1. Therefore the variation of the amplitude is easily linked to the vertical displacements. Another way is to record the variation of the offset of the vertical location of the surface as a function of the drive amplitude. Nevertheless, for the small drive amplitudes, the linear relationship fails and the accuracy is not as good for $A_{\text{free}}$ below 3 to 4 nm.

The experimental approach–retract curves were performed in a gloves box in which the p.p.m. in water was achieved so that capillary forces are negligible. Three different microlevers have been used called OTCL1, OTCL2, OTCL3. The parameters and the samples investigated are given in Table B1.

At a distance of 10 to 20 nm from the surface, dispersion of the value of the phase $\varphi_{\text{free}}$ was sometimes observed as a function of the drive amplitude: $\varphi_{\text{free}} = \langle \varphi_{\text{free}} \rangle \pm 3°$. For the sake of clarity, the phase reported in the figures were all set at the average value.

Finally, the location of the surface being not known, the offset of the $x$-axis is set arbitrarily. When several curves are compared, the offset values were calculated in order to have the first instability of the amplitude and of the phase at the same $x$ location.

[1] F. J. Geissibl, Science **267**, 68 (1995).
[2] S. Kitamura and M. Iwatsuki, Jpn. J. Appl. Phys., Part 1 **35**, 3954 (1996).
[3] Y. Sugarawa, M. Otha, H. Ueyama, and S. Morita, Science **270**, 1646 (1995).
[4] A. Schwarz, W. Allers, U. D. Scwarz, and R. Wiesendanger, Appl. Surf. Sci. **140**, 293 (1999).
[5] R. Lüthi, E. Meyer, L. Howald, H. Haefke, D. Anselmetti, M. Dreier, M. Rüetschi, T. Bonner, R. M. Overney, J. Frommer, and H. J. Güntherodt, J. Vac. Sci. Technol. B **12**, 1673 (1994).
[6] F. J. Geissibl, Phys. Rev. B **56**, 16010 (1997).
[7] J. P. Aimé, R. Boisgard, G. Couturier, and L. Nony, Appl. Surf. Sci. **140**, 333 (1999); J. P. Aimé, R. Boisgard, L. Nony, and G. Couturier, Phys. Rev. Lett. **89**, 3388 (1999).
[8] N. Sasaki and M. Tsukada, Appl. Surf. Sci. **140**, 339 (1999).
[9] S. N. Magonov, V. Elings, and M. H. Wangbo, Surf. Sci. **389**, 201 (1997).
[10] W. Stocker, J. Beckmann, R. Stadler, and J. P. Rabe, Macromolecules **29**, 7502 (1996).
[11] Ph. Leclère, R. Lazzaronni, J. L. Brédas, J. M. Yu, Ph. Dubois, and R. Jérôme, Langmuir **12**, 4317 (1996).
[12] D. Michel, Ph.D. thesis, Université Bordeaux I, 1997.
[13] J. Tamayo and R. Garcia, Appl. Phys. Lett. **71**, 2394 (1997).
[14] L. Wang, Appl. Phys. Lett. **73**, 3781 (1998).
[15] G. D. Haugstad, J. A. Hammerschmidt, and W. L. Gladfelter, in *Microstructures and Tribology of Polymers Surfaces* (ACS, Boston, 1998); G. Haugstad and J. Jones, Ultramicroscopy **76**, 79 (1999); G. Haugstad, Ultramicroscopy (to be published).
[16] J. P. Aimé, D. Michel, R. Boisgard, and L. Nony, Phys. Rev. B **59**, 1829 (1999).
[17] P. Gleyses, P. K. Kuo, and A. C. Boccara, Appl. Phys. Lett. **58**, 2989 (1991); R. Bachelot, P. Gleyses, and A. C. Boccara, Probe Microscopy **1**, 89 (1997).
[18] B. Anczycowsky, D. Krüger, and H. Fuchs, Phys. Rev. B **53**, 15485 (1996).
[19] R. Boisgard, D. Michel, and J. P. Aimé, Surf. Sci. **401**, 199 (1998).
[20] J. Tamayo and R. Garcia, Langmuir **12**, 4430 (1996).
[21] J. de Weger, D. Binks, J. Moleriaar, and W. Water, Phys. Rev. Lett. **76**, 3951 (1996).
[22] R. Boisgard, L. Nony, and J. P. Aimé (unpublished).
[23] J. Israelachvili, *Intermolecular and Surface Forces* (Academic, New York, 1992).
[24] H. Goldstein, *Classical Mechanics* (Addison-Wesley, Reading, 1980).
[25] L. Landau and E. Lifchitz, *Théorie de l'Élasticité*, ed. (MIR, Moscow, 1967).
[26] *Scanning Probe Microscopy in Polymers*, ACS symposium Series 694, edited by Buddy D. Ratner and Vladimir D. Tsukruk (1998), Chap. 16, p. 266.
[27] J. P. Aimé, Z. Elkaakour, C. Odin, T. Bouhacina, D. Michel, J. Curély, and A. Dautant, J. Appl. Phys. **76**, 754 (1994).
[28] S. Gauthier, J. P. Aimé, T. Bouhacina, A. J. Attias, and B. Desbat, Langmuir **12**, 5126 (1996).
[29] C. Fretigny and C. Basire, J. Appl. Phys. **82**(1), 43 (1997).
[30] S. Weigert, M. Dreier, and M. Hegner, Appl. Phys. Lett. **69**, 2834 (1996).